\documentclass[twocolumn,showpacs,preprintnumbers,amsmath,amssymb]{revtex4}

\usepackage{graphicx}
\usepackage{dcolumn}
\usepackage{bm}

\begin{document}

\preprint{APS/123-QED}

\title{Discrete Klein-Gordon models with static kinks free of the Peierls-Nabarro potential}

\author{S. V. Dmitriev$^1$, P. G. Kevrekidis$^2$ and N. Yoshikawa$^1$}
\affiliation{ $^1$ Institute of Industrial Science, the University
of Tokyo, Komaba,
Meguro-ku, Tokyo 153-8505, Japan \\
$^2$ Department of Mathematics and Statistics, University of
Massachusetts, Amherst, MA 01003-4515, USA
 }
\date{\today}

\begin{abstract}
For the nonlinear Klein-Gordon type models, we describe a general
method of discretization in which the static kink can be placed
anywhere with respect to the lattice. These discrete models are therefore
free of the {\it static} Peierls-Nabarro potential.
Previously reported models of this type are
shown to belong to a wider class of models
derived by means of the proposed method. A relevant physical consequence of
our findings is the existence of a wide class of discrete
Klein-Gordon models where slow kinks {\it practically} do not experience
the action of the Peierls-Nabarro potential. Such kinks are not
trapped by the lattice and they can be accelerated by even weak
external fields.
\end{abstract}

\pacs{05.45.-a, 05.45.Yv, 63.20.-e}

\maketitle

\section{Introduction}

Discrete solitons and more specifically
kink-like  topological excitations are ubiquitous structures that
 arise in numerous physical applications ranging from
dislocations or ferroelectric domain
walls in solids, to bubbles in
 DNA, or magnetic chains and Josephson junctions, among
others (see, e.g., \cite{Kivshar} for a recent exposition of
relevant applications). The mobility of such lattice kinks is one
of the key issues in many of these applications, especially since
the pioneering works of \cite{peyrard,yip} which illustrated that
the kinematics on the lattice is dramatically different from the
continuum analog of such equations where constant speed
propagation is typical. Instead, on the discrete substrate, kinks
need to overcome the, so-called, Peierls-Nabarro potential (PNp),
constantly radiating their energy and being eventually trapped by
the lattice. The {\it static} PNp refers to the energy difference
between a stable inter-site centered discrete kink and an
unstable, onsite centered discrete kink. Clearly, as a kink is
travelling from one site to the next, it ``wobbles'' over this
potential energy landscape \cite{boesch}. However, even though
clearly travelling is intimately connected with overcoming the
static PNp without ``radiating'' energy \cite{KW}, this connection
is relatively subtle and the inter-dependence of these two
features (static PNp and travelling) still remains elusive
\cite{flach}. Typically, discrete kinks traveling with finite
velocity have only been obtained for a discrete set of velocities
\cite{yzolo} which makes the motion unstable with respect to
perturbations. There exists a class of more exotic exact solutions
(the so-called ``nanoptera'') where the kink propagates together
with a plane wave having the same velocity \cite{yzolo}.

While the travelling problem is extremely interesting in its own
right, in the present work, we will start by examining the
construction of discrete models with PNp-free kinks, using
a simplified (quasi)static approach. Two classes of discrete
models where static kink can be placed anywhere with respect to
the lattice have been previously derived: one conserving energy
\cite{SpeightKleinGordon} and another one conserving momentum
\cite{PhysicaD}. In both cases the static kink solution can be
obtained from a two-body nonlinear map. In the present paper we
demonstrate that, in general, a discrete version of the first
integral of the static continuum Klein-Gordon field plays the role
of this nonlinear map. Thus we derive a wide class of such models
including the two above-mentioned classes as special cases. The
advantage of this approach is that the kinks
are no longer (typically) trapped by the lattice. Instead they can be
accelerated by even weak external fields. However, a note of caution
should be added here. While one might naively expect that such
solutions
would be intimately connected with slow travelling, it has been
demonstrated numerically that travelling solutions (when they can
be found as e.g. in \cite{yzolo,karpan} for Klein-Gordon lattices,
using the methods of \cite{flesh})
have a sharp lower bound in their wave speed \cite{aigner}. The
existence
of such a threshold illustrates the fact that one should be
particularly
careful in trying to infer features of the travelling problem from
such ``static'' considerations. On the other hand, as the recent
work of Barashenkov, Oxtoby and Pelinovsky demonstrates \cite{dima},
discretizations without PNp are much more natural candidates for
possessing travelling solutions for a isolated wave speeds
(not close to zero).

The presentation of our results will be structured as follows.
Section II will contain the setup and notations used for the
Klein-Gordon models. Section III will present the general
methodology for obtaining static
PNp-free discretizations. Section IV will illustrate the connection
to previously reported models. Section V will focus on the special
case example of the $\phi^4$ model, for which our numerical observations
will be presented in section VI. Finally in section VII, we will
summarize our findings and present our conclusions.

\section{Setup}

We consider the Lagrangian of the Klein-Gordon field,
\begin{equation}
L =\int_{-\infty}^{\infty} \left[
\frac{1}{2}\phi_t^2-\frac{1}{2}\phi_x^2-V(\phi)\right]dx\,,
\label{KleinGordonHam}
\end{equation}
and the corresponding equation of motion,
\begin{equation}
\phi _{tt}  = \phi _{xx}  - V'(\phi)\equiv D(x)\,.
\label{KleinGordon}
\end{equation}

Topological solitons (kinks) are possible only if $V(\phi)$ has at
least two minima $\phi_{01}$ and $\phi_{02}$, where
$V^{\prime}(\phi_{0i})=0$ and $V^{\prime\prime}(\phi_{0i})>0$.
Obviously, $\phi=\phi_{01}$ and $\phi=\phi_{02}$ are the
stationary solutions to Eq. (\ref{KleinGordon}). We will study the
properties of kinks that interpolate between these two stationary
solutions. Our considerations allow one to treat the cases when
other minima appear in between the two minima, $\phi_{01}$ and
$\phi_{02}$, connected by the kink.

Equation (\ref{KleinGordon}) will be discretized on the lattice
$x=nh$, where $n=0,\pm 1, \pm 2 ...$, and $h$ is the lattice
spacing.

For brevity, when possible, we will use the notations
\begin{equation}
\phi _{n - 1}  \equiv l,\,\,\,\,\,\phi _n  \equiv m,\,\,\,\,\,\phi
_{n + 1}  \equiv r\,. \label{Notation}
\end{equation}

We would like to construct a nearest-neighbor discrete analog to
Eq. (\ref{KleinGordon}) of the form
\begin{equation}
\ddot{m} = D(C,l,m,r), \label{KleinGordonDiscrete}
\end{equation}
where $C>0$ is a parameter related to the lattice spacing $h$ as
$C=1/h^2$, such that in the continuum limit $(C\rightarrow \infty)$,
$D(C,l,m,r) \rightarrow D(x)=\phi_{xx}- V'(\phi)$.

Note that in this context, the ``standard'' discretization emerges
in the form: $D(C,l,m,r)=C(l-2m+r)-V'(m)$. Generalizations of this
model will be discussed in the form
\begin{equation}
\ddot{m}= C(l-2m+r) - B(l,m,r), \label{KleinGordonDiscrS}
\end{equation}
where $B(l,m,r)$ has $V'(\phi)$ as the continuum limit.

We will characterize a model as PNp-free if a {\em static} kink
can be placed anywhere with respect to the lattice (continuum,
rather than discrete, set of equilibrium solutions). This is
equivalent to demanding that the kink must have an neutral
direction, or (from Noether's theory \cite{arnold}) a Goldstone
translational mode. It is natural to categorize this definition of
PNp-free model as ``static'' or ``quasi-static'', in the sense
that it does not involve the kinematic or dynamical properties of the model.

On the other hand, one can demand the absence of PNp at finite
kink velocities. This can be transformed to the demand that the
discrete model supports the exact traveling wave solutions and
this demand can be called ``dynamic'' definition; see e.g. \cite{flach}
for such travelling wave examples, where the ``static'' definition
of the PNp clearly fails.

In this paper we aim to construct the models PNp-free in the
static sense as a first (yet nontrivial) step towards
understanding the nature of the discrete travelling problem (see
also the comments above).

We will also focus on the existence of physically motivated
conserved quantities
for the derived models. Hamiltonian models are energy-conserving
models and the models with $dM/dt=0$, where
\begin{eqnarray}
M= \sum_{n=-\infty}^{\infty} \dot{\phi}_n
\left(\phi_{n+1}-\phi_{n-1} \right), \label{mom1}
\end{eqnarray}
will be called momentum-conserving models. As was shown in
\cite{PhysicaD}, the discrete model of Eq.
(\ref{KleinGordonDiscrete}) conserves the momentum of Eq. (\ref{mom1}),
if it can be presented in the form
\begin{eqnarray}
\ddot{m}=\frac{{\cal H}(m,r)-{\cal H}(l,m)}{r-l}. \label{mom2}
\end{eqnarray}
This can be verified by calculating
\begin{eqnarray}
\frac{dM}{dt}=\sum_n \ddot{\phi}_n
(\phi_{n+1}-\phi_{n-1})\nonumber \\
= \sum_n [{\cal H}(\phi_{n},\phi_{n+1}) - {\cal
H}(\phi_{n-1},\phi_{n})]=0, \label{mom3}
\end{eqnarray}
where we have used the fact that the terms
$\dot{\phi}_n(\dot{\phi}_{n+1}-\dot{\phi}_{n-1})$ cancel out due
to telescopic summation.

\section{Static PNp-free discretization}

Our aim here will be to discretize Eq. (\ref{KleinGordon}) in a
symmetric way, so that the static kink solution can be found from
a reduced first-order difference equation. According to
\cite{SpeightKleinGordon}, if we achieve that, then we are going
to have a one-parameter family of solutions with the possibility
to place equilibrium kinks anywhere with respect to the lattice
(and hence, PNp-free in the static sense).

The first integral of the steady state problem in Eq.
(\ref{KleinGordon}), $\phi_x - \sqrt{2V(\phi)} = 0$ (with zero
integration constant), can be written in the form
\begin{equation}
w(x) \equiv g(\phi_x) - g\left(\sqrt{2V(\phi)}\right) = 0 \,,
\label{StaticFirstIntegral2}
\end{equation}
where $g$ is a continuous function.

Our plan will then be the following:
\begin{itemize}
\item discretize the first-order differential equation
of Eq. (\ref{StaticFirstIntegral2}) using a first order difference
scheme $w(l,m)=0$.
\item Then express the right-hand side of Eq.
(\ref{KleinGordon}) as a sum of terms containing derivatives,
e.g., $dw/dx$, $dw/d\phi$, etc.
\item As a result, discretizations of such terms,
e.g., $dw/dx \sim \sqrt{C}[w(m,r)-w(l,m)]$, vanish for $w(l,m)=0$
(or otherwise stated: the construction of the equilibrium solution
is converted to a first order difference problem).
Then, the static kink (PNp-free, by construction)
solutions for the obtained discrete model
can be found from this two-site problem.
\end{itemize}

In the following, we will consider a particular case of Eq.
(\ref{StaticFirstIntegral2}) with $g(\xi)=\xi^2$, for which we
introduce the notation
\begin{eqnarray}
u(x) \equiv \phi_x^2 - 2V(\phi) = 0 \,, \label{ucont1}
\end{eqnarray}
and the following two-site discrete analog
\begin{eqnarray}
u(l,m) \equiv C(m-l)^2 - 2V(l,m) = 0 \,. \label{udisc1}
\end{eqnarray}
We will also use the shorthand notations,
\begin{equation}
u_l=u(l,m)\,,\,\,\,\,\,\,\,\,\,\,\,\,u_m=u(m,r). \label{Shortnot}
\end{equation}

We have assumed that the Klein-Gordon field supports kink
solutions. Then, at least for the case of weak discreteness, Eq.
(\ref{udisc1}) also supports static kinks because it is nothing
but a discretization of the first integral of static version of
Eq. (\ref{KleinGordon}) (see also \cite{SpeightKleinGordon}).

The next step is then to find a discretization of the right-hand side of
Eq. (\ref{KleinGordon}), $D(x)$, which vanishes when Eq. (\ref{udisc1})
is fulfilled.

One simple possibility comes from the following finite difference
\begin{eqnarray}
D_1(l,m,r) \equiv  \frac{u_m-u_l}{r-l} \rightarrow
\frac{1}{2}\frac{du}{d\phi}= D(x). \label{r1}
\end{eqnarray}

One can also consider, more generally, continuous functions
$q(u_l,h)$ such that
$q(0,h)=0$ and, in the continuum limit, $q(u,0)=u$ and
$\frac{dq}{du}(u,0)=1$. For example, one can take $q=(e^{hu}-1)/h$
or $q=u+\sum_{n>1}A_nh^{n-1}u^n$ with constant $A_n$, etc. Then,
\begin{equation}
\frac{1}{2}\frac{dq}{d\phi}\left(\frac{dq}{du}\right)^{-1} = D(x).
\label{Dr2}
\end{equation}
Discretizing the left-hand side of Eq. (\ref{Dr2}) we obtain
\begin{equation}
D_2 = \frac{1}{2}\frac{q(u_m,h) - q(u_l,h)}{r-l}\left[
\frac{1}{q'(u_l)} + \frac{1}{q'(u_m)} \right]. \label{r2}
\end{equation}

Inspired by \cite{SpeightKleinGordon}, we note that, in
the continuum limit,
\begin{eqnarray}
\frac{v(m,r)}{r-m}-\frac{v(l,m)}{m-l} \rightarrow
\frac{dv}{d\phi}-v\frac{\phi_{xx}}{\phi_x^2}\,, \label{ContLim}
\end{eqnarray}
and find
\begin{eqnarray}
D_3 \equiv \frac{u_m}{r-m}-\frac{u_l}{m-l} + \sqrt{2V(l,m,r)}\times \nonumber \\
\left(\frac{ \sqrt{C(r - m)^2-u_m}}{r-m}-\frac{ \sqrt{C(m -
l)^2-u_l}}{m-l}\right) \nonumber \\
\rightarrow \frac{du}{d\phi}-u\frac{\phi_{xx}}{\phi_x^2} +
\sqrt{2V}\left(\frac{d\sqrt{2V}}{d\phi}-\sqrt{2V}\frac{\phi_{xx}}{\phi_x^2}\right)
\nonumber \\ = D(x). \label{r3}
\end{eqnarray}

Since the expressions for $D_i(l,m,r)$ given by Eqs. (\ref{r1}),(\ref{r2}) and
(\ref{r3}) tend to $D(x)$ in the continuum limit, one can write the
following discrete analog to the Klein-Gordon equation Eq.
(\ref{KleinGordon})
\begin{equation}
\ddot{m}  = \sum_i b_iD_i(l,m,r), \,\,\,\,\,\,\,\,\,{\rm
where}\,\,\,\,\,\,\,\,\,\sum_ib_i=1. \label{KleinGordonDiscr}
\end{equation}
Then, by construction, any structure derived from the two-site problem of
Eq. (\ref{udisc1}) is a static solution of Eq.
(\ref{KleinGordonDiscr}) and hence, the latter is the static
PNp-free discrete model.

The model of Eq. (\ref{KleinGordonDiscr}) can be generalized in a
number of ways. For example, function $D_3$, Eq. (\ref{r3}), can
be modified choosing different functions $V(l,m,r)$ to discretize
$V(\phi)$. Then, the modified $\tilde{D}_3$ can be added to the
linear combination in the right-hand side of Eq.
(\ref{KleinGordonDiscr}).

The model of Eq. (\ref{KleinGordonDiscr}) can be also generalized by
appending terms which disappear in the continuum limit
 and  ones that
vanish upon substituting $u_l=0$ and $u_m=0$. For example, the
derivative $df(u)/d\phi$ can be discretized as
$2[f(u_m)-f(u_l)]/(r-l)$ or as $2f'(u_l/2+u_m/2)(u_m-u_l)/(r-l)$.
If then we have difference of such terms in the equation of
motion, then in the continuum limit they will cancel out.

Any term in the right-hand side of Eq. (\ref{KleinGordonDiscr})
can be further modified by multiplying by a continuous function
$e(C,l,m,r)$, whose continuum limit is unity
(see e.g. \cite{Saxena} for such an example, also discussed
in more detail below).

Generally speaking, the discrete PNp-free Klein-Gordon models
derived here do not conserve either an energy, or a
momentum-like quantity. However, as it will
be demonstrated below, they contain energy-conserving and
momentum-conserving subclasses.

\section{Connection with Previously Reported Models}

One energy-conserving PNp-free
Klein-Gordon model has been derived by Speight with co-workers
\cite{SpeightKleinGordon} with the use of the Bogomol'nyi argument
\cite{Bogom}. Their model, can be written in the form of Eq.
(\ref{KleinGordonDiscrS}), with the Lagrangian
\begin{eqnarray}
L = \frac{1}{2}\sum\limits_n \dot \phi_n^2
-\frac{C}{2}\sum\limits_n\left( {\phi_n - \phi_{n-1} } \right)^2  \nonumber \\
- \sum\limits_n\left({\frac{{G(\phi_n) - G(\phi_{n-1})}}{{\phi_n -
\phi_{n-1}}}} \right)^2,  \nonumber \\
{\rm where}\,\,\,\,\,\,\,\,\,G^{\prime}(\phi) = \sqrt{V(\phi)}.
\label{SpeightHam}
\end{eqnarray}
The static kink solution can then be derived from the lattice Bogomol'nyi
equation \cite{SpeightKleinGordon}, which can be taken in the form
\begin{eqnarray}
U(l,m) = C(m - l)^2 - 2\left(\frac{G(m) - G(l)}{m - l}\right)^2 =
0, \label{Speight1}
\end{eqnarray}
which is a particular case of Eq. (\ref{udisc1}). The equation of
motion derived from Eq. (\ref{SpeightHam}), written in terms of
Eq. (\ref{Speight1}), is
\begin{eqnarray}
\ddot{m}= \frac{U_m}{r-m}-\frac{U_l}{m-l} +\sqrt{2V(m)} \times \nonumber \\
\left(\frac{ \sqrt{C(r - m)^2-U_m}}{r-m}-\frac{ \sqrt{C(m -
l)^2-U_l}}{m-l}\right). \label{Speight3}
\end{eqnarray}
The right-hand side of Eq. (\ref{Speight3}) is a particular case
of $D_3(l,m,r)$ given by Eq. (\ref{r3}) with $V(l,m,r)=V(m)$.

Momentum-conserving PNp-free models were proposed in
\cite{PhysicaD} and further studied in \cite{Submitted}. They are
the non-Hamiltonian models of the form
\begin{eqnarray}
\ddot{m}= D_1 (l,m,r), \label{PhysD3}
\end{eqnarray}
where $D_1$ is given by Eq. (\ref{r1}). Notice that Eq.
(\ref{PhysD3}) can be mapped into the formulation
 of Eq. (\ref{mom2}). Static kink
solutions in this model can be found from Eq. (\ref{udisc1}).

If Eq. (\ref{udisc1}) is taken in the particular form of Eq.
(\ref{Speight1}), then the momentum-conserving PNp-free model Eq.
(\ref{PhysD3}) and the energy-conserving PNp-free model Eq.
(\ref{Speight3}) have exactly the same static kink solutions. It
has been proved that a standard nearest-neighbor discrete
Klein-Gordon model conserving both energy and momentum does not
exist \cite{Submitted}.

\section{Application to the $\phi^4$ model}

As an example, we will
discretize the well-known $\phi^4$ field theory with the potential
\begin{equation}
V(\phi) = \frac{1}{4}\left(1-\phi^2\right)^2\,.
\label{Phi4potential}
\end{equation}

By construction, the PNp-free models derived above are written in
singular form. In this form the equations are inconvenient in
practical simulations and one may wish to find such particular
cases when singularities disappear.

For example, for the energy-conserving PNp-free model expressed by
Eqs. (\ref{SpeightHam})-(\ref{Speight3}), singularity always
disappears when $G(\phi)$ is polynomial \cite{SpeightKleinGordon}.
Particularly, for the $\phi^4$ model with the potential Eq.
(\ref{Phi4potential}), one obtains from Eq. (\ref{Speight3}) the
following energy-conserving PNp-free discretization derived in
\cite{SpeightKleinGordon}
\begin{eqnarray}
\ddot m = \left(C + \frac{1} {6} \right)(l + r
- 2m) + m \nonumber \\
- \frac{1} {{18}}\left[ {2m^3  + (m + l)^3 + (m + r)^3 } \right],
\label{SpeightPhi4}
\end{eqnarray}
whose static kink solution can be found from Eq. (\ref{Speight1}),
which, for the $\phi^4$ potential, obtains the form
\begin{eqnarray}
3\sqrt{2C}(m - l) + m^2 + lm + l^2-3 = 0. \label{SpeightPhi4kink}
\end{eqnarray}

Now let us turn to the momentum-conserving model. Substituting Eq.
(\ref{udisc1}) into Eq. (\ref{PhysD3}) we obtain
\begin{eqnarray}
\ddot m = C(r-2m+l) -2\frac{V(m,r)-V(l,m)}{r-l}.
\label{Rphi4momcons}
\end{eqnarray}
To remove the singularity, $V(l,m)$ should be taken in the
symmetric form $V(l,m)=V(m,l)$, e.g., as
\begin{eqnarray}
V(l,m)=(1/4) - (\alpha/2)(m^2+l^2)+(\alpha -1/2 ) ml \nonumber \\
+ (\beta/2) \left( {m^3  + l^3 } \right) - (\beta/2) ml\left( {m + l} \right) \nonumber \\
+ (\gamma/2) \left( {m^4  + l^4 } \right) + (\delta/2) ml\left( {m^2  + l^2 } \right) \nonumber \\
- \left( { \gamma + \delta -1/4} \right)m^2 l^2,
\label{HbasicPhi4Transformed}
\end{eqnarray}
with free parameters $\alpha$, $\beta$, $\gamma$, and $\delta$. In
the continuum limit, when $l\rightarrow m$ and $r\rightarrow m$,
Eq. (\ref{HbasicPhi4Transformed}) reduces to $V(\phi)$.
Substituting Eq. (\ref{HbasicPhi4Transformed}) into Eq.
(\ref{Rphi4momcons}) we obtain the following momentum-conserving
PNp-free $\phi^4$ model derived in \cite{Submitted}
\begin{eqnarray}
\ddot m = \left( {C + \alpha } \right)(l - 2m+ r )
+ m \nonumber \\
-\beta (l^2  + lr + r^2 ) + \beta m(l + r + m) \nonumber \\
-\gamma (l^3 + r^3  + l^2 r + lr^2 ) - \delta m(l^2  + m^2  + r^2
+ lr)\nonumber \\
+ ( 2\gamma  + 2\delta -1/2)m^2 (l + r). \label{PhysicaDPhi4}
\end{eqnarray}

The momentum-conserving model Eq. (\ref{PhysicaDPhi4}) with
$\alpha=\beta=\gamma=\delta=0$ can be written in the form
\begin{eqnarray}
\ddot m = \left(1-\frac{m^2}{2C}\right)C(l - 2m+ r )+ m -m^3.
\label{Saxena1}
\end{eqnarray}
The following energy-conserving model, studied in \cite{Saxena},
\begin{equation}
\ddot m = C(l - 2m+ r )+ \frac{m-m^3}{1-m^2/(2C)}, \label{Saxena2}
\end{equation}
has the same continuum limit as model Eq. (\ref{Saxena1}).
Furthermore, it can be derived from Eq. (\ref{Saxena1}) by
multiplication with a factor $e(C,l,m,r)=1/(1-m^2/(2C))$, which
possesses
a unit continuum limit. The model Eq. (\ref{Saxena1}) is PNp-free and thus,
model Eq. (\ref{Saxena2}) is also PNp-free since they have the
same static solutions derivable from $C(m-l)^2-(1-ml)^2/2=0$.
Thus, we have another example when energy-conserving and
momentum-conserving PNp-free models have exactly the same static
kink solutions.

It is interesting to note that the energy-conserving model of Eq.
(\ref{Saxena2}) cannot be constructed by the method reported in
\cite{SpeightKleinGordon} where discretization of the anharmonic
term always involves $\phi_{n-1}$ and $\phi_{n+1}$. More generally
than it is done in \cite{SpeightKleinGordon}, the problem of
finding the energy-conserving PNp-free models can be formulated as
follows. We need to discretize the potential energy of the
Lagrangian Eq. (\ref{KleinGordonHam}) in a way that the
corresponding equation of static equilibrium is satisfied when Eq.
(\ref{StaticFirstIntegral2}) is satisfied. Both energy-conserving
models discussed above are the solutions of this problem.

As an example of model conserving neither energy, nor momentum we take
Eq. (\ref{r2}) for the case of $q(u,h)=u+Ahu^2$ with constant $A$
and obtain
\begin{equation}
\ddot{m}=\frac{u_m-u_l}{r-l}\frac{(1+Ahu_l+Ahu_m)^2} {(1+2Ahu_l)
(1+2Ahu_m)}. \label{Xr2}
\end{equation}
This model can be obtained from the momentum-conserving model
defined by Eq. (\ref{r1}) by multiplying by another function that
reduces to unity in the continuum limit ($h \rightarrow 0$).
Obviously, the original momentum-conserving model and model Eq.
(\ref{Xr2}) have the same static kink solutions. It can be
demonstrated that these two models also have the same spectra of
small amplitude vibrations and the same frequencies of kink
internal modes.

\section{Numerics}

In our recent work \cite{Submitted}, some
properties of kinks were compared for the ``standard''
energy-conserving $\phi^4$ discretization having PNp,
\begin{eqnarray}
\ddot{m}=C(l + r - 2m)+m -m^3, \label{PHI4Classic}
\end{eqnarray}
with the PNp-free models conserving energy Eq. (\ref{SpeightPhi4})
and momentum Eq. (\ref{Saxena1}).

It was found that the mobility of kinks in the PNp-free models is
higher and also that in the momentum-conserving, PNp-free models
a kink self-acceleration effect may be observed. The origin of the
effect is the non-conservative (non self-adjoint)
nature of the model which, however,
can be noticed only for asymmetric trajectories of particles when
kink passes by \cite{Submitted}. If the trajectories are
symmetric, there is no energy exchange with the surroundings and
kink dynamics is the same as in energy-conserving models, e.g., the
kink self-acceleration effect disappears. Kinks in some of the
momentum-conserving models was found to have internal modes with
frequencies above the phonon spectrum. Such modes do not radiate
and they can have large amplitudes storing a considerable amount of
energy.

Here we present/compare results for the energy-conserving
PNp-free model Eq. (\ref{Saxena2}) and the PNp-free model of Eq.
(\ref{Xr2}), generally speaking, conserving neither energy nor momentum.
For the latter model we take $u_l$ in the form of Eq.
(\ref{udisc1}) where the $\phi^4$ potential is discretized
according to Eq. (\ref{HbasicPhi4Transformed}) and, for the sake
of simplicity, we set $\alpha=\beta=\gamma=\delta=0$. We obtain
\begin{eqnarray}
\ddot m = \left[\left(1-\frac{m^2}{2C}\right)C(l - 2m+ r )+ m
-m^3\right]\times\nonumber \\
\frac{(1+Ahu_l+Ahu_m)^2} {(1+2Ahu_l) (1+2Ahu_m)}\,,\nonumber \\
{\rm where}\,\,\,\,\,\,\,u_l=C(m-l)^2-(1-ml)^2/2\,.
\label{Noconservation}
\end{eqnarray}
For $A=0$, Eq. (\ref{Noconservation}) coincides with the
momentum-conserving model Eq. (\ref{Saxena1}).

In the model Eq. (\ref{Noconservation}), the static kink
solutions, phonon spectra, and frequencies of kink internal modes
are $A$-independent. The energy-conserving model Eq.
(\ref{Saxena2}) has the same static kink solutions as model Eq.
(\ref{Noconservation}) but their spectra are different. The linear
vibration spectrum of the vacuum for Eq. (\ref{Noconservation}) is
$\omega^2=2+(4C-2)\sin^2(\kappa/2)$ and that for Eq.
(\ref{Saxena2}) is $\omega^2=4C/(2C-1)+4C\sin^2(\kappa/2)$, while
the one for the classical model Eq. (\ref{PHI4Classic}) is
$\omega^2=2+4C\sin^2(\kappa/2)$.

\begin{figure}
\includegraphics{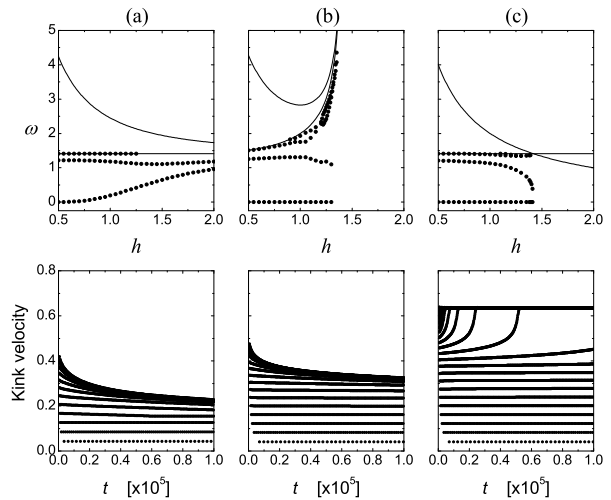}
\caption{Upper panels: boundaries of the linear spectrum of the
vacuum (solid lines) and kink internal mode frequencies (dots) as
functions of the lattice spacing $h=1/\sqrt{C}$. Lower panels:
time evolution of kink velocity for different initial velocities
and $h=0.7$. The results are shown for (a) classical $\phi^4$
model, Eq. (\ref{PHI4Classic}), (b) PNp-free model conserving
energy, Eq. (\ref{Saxena2}), and (c) PNp-free model conserving
momentum, Eq. (\ref{Noconservation}) at $A=0$.} \label{Figure1}
\end{figure}

\begin{figure}
\includegraphics{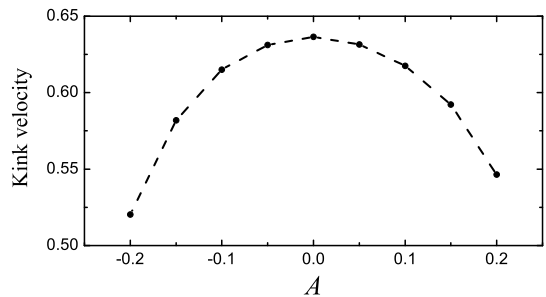}
\caption{The kink velocity in the regime of steady motion [see bottom
panel in Fig. \ref{Figure1} (c)] for the PNp-free $\phi^4$ model
Eq. (\ref{Noconservation}) is shown as
a function of parameter $A$. For $|A| > 0.2$, the
kink self-acceleration effect disappears.}
\label{Figure2}
\end{figure}

The top panels of Fig. \ref{Figure1} present the boundaries of the
linear vibration spectrum of the vacuum (solid lines) and the kink
internal modes (dots) as the functions of lattice spacing $h$ for
(a) the classical $\phi^4$ model of Eq. (\ref{PHI4Classic}), (b)
the PNp-free model of Eq. (\ref{Saxena2}) conserving energy, and
(c) the PNp-free model of Eq. (\ref{Noconservation}) at $A=0$
conserving momentum. In PNp-free models kinks possess a zero
frequency, Goldstone translational mode. Since all models
presented in Fig. \ref{Figure1} share the same continuum $\phi^4$
limit, their spectra are very close for small $h(<0.5)$.

The bottom panels of Fig. \ref{Figure1} show the time evolution of
kink velocity for the corresponding models at $h=0.7$ for kinks
launched with different initial velocities. To boost the kink we
used the semi-analytical solution for the normalized Goldstone
mode, whose amplitude serves as a measure of the initial kink
velocity. One can see that the mobility of kinks in the PNp-free
models shown in (b) and (c) is higher than in the classical model
having PNp and shown in (a). In the energy-conserving models shown
in (a) and (b), the kink velocity decreases monotonically due to the
energy radiation. Non-Hamiltonian momentum-conserving model in (c)
shows the effect of kink self-acceleration discussed in
\cite{Submitted}.

It is interesting to study what happens when the parameter $A$ in Eq.
(\ref{Noconservation}) deviates from zero and the conservation law
of the model (momentum conservation) disappears.
We found that the effect of kink
self-acceleration, which can be seen in the bottom panel of Fig.
\ref{Figure1} (c) for $A=0$, remains for $|A|<0.2$ but the value
of the kink velocity in the steady motion regime decreases with
increase in $|A|$ as it is presented in Fig. \ref{Figure2}. For
$|A|>0.2$ kink self-acceleration effect disappears and kink
velocity gradually decreases with time. From the above, we infer that
properties such as the self-acceleration (for momentum-conserving
models) or the Bogomol'nyi bounds (for energy-conserving discretizations)
render such models rather special within the
broader class of PNp free models. However, the critical ingredient for
the more general feature of ({\it static}) PN absence exists in the form of a
reduction
of the second order problem into a first order.

\section{Conclusions}

A general procedure for deriving discrete
Klein-Gordon models whose static kinks can be placed anywhere with
respect to the underlying
lattice was described. Such models are called {\em
static} PNp-free models. It was demonstrated that the models of
this kind derived earlier
\cite{SpeightKleinGordon,PhysicaD,Submitted,Saxena} are
special cases of the wider family of models derived here.

Static kink solutions for the PNp-free models can be found
from the nonlinear algebraic equation of the form $u(l,m)=0$,
which is a discrete analog of the first integral of the static
continuum Klein-Gordon equation of motion. This ensures the
existence of static kink solutions at least for the regime of
sufficiently weak discreteness and smooth background potential.
The range of the discreteness parameter supporting stable static kinks
varies according to the specific properties  of the model.

In this paper we have discussed only nearest-neighbor
discretizations. However, one can easily write down a PNp-free
model involving more distant neighbors by replacing Eq.
(\ref{KleinGordonDiscr}) with higher-order finite difference
operators approximating Eq. (\ref{KleinGordon}), keeping the
two-point approximation, Eq. (\ref{udisc1}), for the first
integral of Eq. (\ref{ucont1}).

Discrete kinks in the static PNp-free models possess the
zero-frequency translational Goldstone mode and they can (almost) freely
move with at least infinitesimally small velocity. Such kinks are
not trapped by the lattice and they can be accelerated by even
weak external fields.

As a topic for future studies, it would be interesting to find any
possible relation between models constructed here and models that
support traveling kink solutions for finite kink velocity. Such
connections are apparently under intense investigation \cite{dima}
and should provide a framework for understanding travelling in
dispersive lattice systems.



\end{document}